# Gray spatial solitons in nonlocal nonlinear media

Yaroslav V. Kartashov and Lluis Torner

*ICFO-Institut de Ciencies Fotoniques, Mediterranean Technology Park, and Universitat Politecnica de Catalunya, 08860 Castelldefels (Barcelona), Spain*

We study gray solitons in nonlocal nonlinear media and show that they are stable and can form bound states. We reveal that gray soliton velocity depends on the nonlocality degree, and that it can be drastically reduced in highly nonlocal media. This is in contrast to the case of local media, where the maximal velocity is dictated solely by the asymptotic soliton amplitude.

Nonlocality is ubiquitous in several nonlinear optical materials, such as nematic liquid crystals, thermo-optical materials, plasmas, and photorefractive crystals. It has been shown to profoundly affect beam interactions [1]. Thus, interaction between one-dimensional bright solitons in nonlocal media depends on spatial separation [2,3] so that bound states of bright solitons may form [4-6]. Nonlocal materials sustain complex two-dimensional solitons [7-16], while dark solitons may exist in defocusing nonlocal media. Strongly nonlocal response can cause attraction between two dark solitons [17,18] which always repel each other in local medium. A characteristic signature of dark solitons in nonlocal media is the presence of non-monotonic tails affording the formation of bound states [17].

Dark solitons, however, are a particular type of gray solitons having antisymmetric phase profiles, but with a smaller and more gradual phase shift [19]. Gray solitons move in the transverse plane upon propagation, with a grayness that depends on the soliton velocity. In this Letter we introduce properties of gray solitons in nonlocal nonlinear media. We show that nonlocal gray solitons are stable and can form bound states. We discover that nonlocality imposes drastic restrictions on gray soliton velocity.

Our theoretical model is based on two coupled equations for the light field amplitude $q$ and nonlinear contribution to refractive index $n$ describing the propagation of laser beam along the $\xi$ axis of a nonlocal defocusing medium:

$$i\frac{\partial q}{\partial \xi} = -\frac{1}{2}\frac{\partial^2 q}{\partial \eta^2} - qn,$$
$$d\frac{\partial^2 n}{\partial \eta^2} - n = |q|^2, \qquad (1)$$

where $\eta$ and $\xi$ stand for the transverse and longitudinal coordinates scaled to the characteristic width of the dark notch of the soliton profile and to the diffraction length, respectively. When nonlocality parameter $d \to 0$ one recovers the local case, while $d \to \infty$ corresponds to a strong nonlocality. The nonlinear contribution to refractive index is given by $n = -\int_{-\infty}^{\infty} G(\eta - \lambda)|q(\lambda,\xi)|^2 \, d\lambda$, where $G(\eta) = (1/2d^{1/2})\exp(-|\eta|/d^{1/2})$ is the response function of the nonlocal medium. Defocusing nonlocal nonlinearities that allow formation of dark and gray solitons are encountered in various physical systems, including atomic vapors, weakly absorbing liquids [18], and liquid crystals featuring a thermal response [20]. While the details of the response function may be different in each case, the model (1) is expected to capture the general physics of nonlocal gray solitons. Note that Eq. (1) can be simplified in the limiting cases of strong and weak nonlocality. Thus, at $d \to 0$ the equation for $n$ can be solved at the first order of perturbation theory to give $n = -(1 + d\partial^2/\partial \eta^2)|q|^2$, a result that allows obtaining corrections to the parameters of local gray solitons in analytic form [19]. At $d \to \infty$ one can use the fact that the response function width is much larger than the dark soliton notch width to simplify the expression for $n$ and to obtain a simple equation for $q$ by using expansions for $G(\eta)$ function.

We searched for gray solitons of Eqs. (1) numerically in the moving coordinate frame $\zeta = \eta - \alpha\xi$, with the form $q(\zeta,\xi) = [u(\zeta) + iv(\zeta)]\exp(ib\xi)$, where $u$ and $v$ are the real and imaginary field amplitudes, and $b$ is the propagation constant. In the case of single solitons we assume that $u(-\zeta) = -u(\zeta)$ and $v(-\zeta) = v(\zeta)$. At $d = 0$ Eqs. (1) have the analytical solution $u = (-b - \alpha^2)^{1/2} \tanh[(-b - \alpha^2)^{1/2}\zeta]$, and $v = \alpha$, that describes a gray soliton moving with velocity $\alpha$, with $\alpha^2 < -b$. The propagation constant $b$ dictates the asymptotic values of soliton intensity $|q(\eta \to \pm\infty, \xi)|^2 = -b$ and refractive index $n(\eta \to \pm\infty) = b$, while the velocity $\alpha$ sets the soliton grayness, defined as $g = \min|q|^2 = \alpha^2$. We analyze the impact of soliton velocity $\alpha$ and nonlocality degree $d$ on the properties of solitons with equal asymptotic intensities at $\eta \to \pm\infty$ and initially set $b = -1$ (see below for other values). It is useful to introduce the rescaled energy flow $U_r$ and momentum $P_r$:

$$U_r = \int_{-\infty}^{\infty} \left| -b - |q|^2 \right| d\eta,$$
$$P_r = \frac{i}{2} \int_{-\infty}^{\infty} \left( q \frac{\partial q^*}{\partial \eta} - q^* \frac{\partial q}{\partial \eta} \right) \left| 1 + \frac{b}{|q|^2} \right| d\eta. \tag{2}$$

The nonlocality drastically affects gray soliton properties. Representative examples of intensity profiles $w = u^2 + v^2$ of single gray solitons are depicted in Figs. 1(a) and 1(b). The asymptotic intensity and refractive indices for nonlocal gray solitons at $\eta \to \pm\infty$ coincide with those in local media. The soliton grayness $g = \min|q|^2$ (minimal intensity value) monotonically increases with $\alpha$, but in contrast to gray solitons in local media that feature smooth shapes, rapidly moving nonlocal gray solitons develop multiple intensity oscillations around the main intensity deep (Fig. 1(b)). These oscillations become more pronounced for high nonlocality degrees. They exist even at $\alpha \to 0$, though the number of oscillations increases while their amplitude decreases with $\alpha$. The transverse phase distributions (Fig. 2) confirm that solitons from Figs. 1(a) and 1(b) correspond to single gray solitons. The development of shape oscillations is accompanied by divergence of $U_r$ which turns out to be a nonmonotonic function of $\alpha$ (Fig. 3(a)), in contrast to the case of local media where energy $U_r = 2(-b-\alpha^2)^{1/2}$ monotonically decreases with $\alpha$ and vanishes at $\alpha \to -b = 1$. At $\alpha = 0$ and $d > 0$ the rescaled energy $U_r$ exceeds the value $2(-b)^{1/2}$ of energy in local media because nonlocality generates oscillations even on dark solitons.

There is a maximal soliton velocity, $\alpha_m$, at which the gray soliton vanishes completely, since $g(\alpha \to \alpha_m) = 1$. The maximal soliton velocity in nonlocal media is lower than that in local media (Fig. 3(b)). Therefore, nonlocality of nonlinear response strongly affects the mobility of gray solitons. Figure 4(a) show how the maximal soliton velocity decreases with $d$. The soliton grayness is a monotonically increasing function of the nonlocality degree for a fixed velocity $\alpha$, and it tends to unity when $d$ approaches a certain limiting value (Fig. 4(b)). Notice that in contrast to $U_r$, the rescaled momentum $P_r$ is a monotonically increasing function of $\alpha$, which in accordance with stability theory for gray solitons [19] might be considered as an indication of soliton stability.

We found that solitons exhibit qualitatively similar properties for all values of $b$, i.e. for each $b$ there exist maximal velocity that monotonically decreases with $d$. Increasing $|b|$ results in a monotonic increase of the maximal possible soliton velocity at fixed $d$. Thus, at $d = 5$ the maximal velocity is $\alpha_{\mathrm{m}} = 0.631$ for $b = -1$, while one gets $\alpha_{\mathrm{m}} = 0.758$ for $b = -2$, and $\alpha_{\mathrm{m}} = 0.913$ for $b = -4$. Therefore, in nonlocal media the maximal velocity increases with $|b|$ slower than in local media, where $\alpha_{\mathrm{m}} = (-b)^{1/2}$.

To study gray soliton stability we performed simulations of Eqs. (1) with input conditions $q|_{\xi=0} = (u + iv)(1 + \rho)$, where $\rho(\eta) \ll 1$ is a broadband noise. Since we used a split-step Fourier method in simulations, we employed combinations of two identical solitons with opposite signs of $u$ separated by a huge distance. We found that single gray solitons are stable, including slowly moving almost dark solitons and solitons moving with velocities close to $\alpha_{\mathrm{m}}$ and featuring multiple oscillations (Figs. 5(a) and 5(b)). We studied the propagation of gray solitons nested on a wide Gaussian envelope, which does not qualitatively affect soliton dynamics. Only when, after a long-distance propagation, the moving gray soliton reaches the region where the background intensity of the host beam starts to decrease slowly, the soliton slightly accelerates, similarly to gray soliton propagating on finite background in local nonlinear media [19,21,22].

The presence of nonmonotonic tails in single gray solitons suggest the possibility to form bound states of several gray solitons moving with the same velocity. The simplest bound state of two gray solitons is shown in Figs. 1(c) and 1(d). Its intensity distribution possesses local maximum in between two gray solitons forming the bound state, while the refractive index profile has a shape of a waveguide which is capable to keep the complex gray solitons together. A multiple intensity oscillations develop in the profile of the bound states when their velocity approaches the maximal value $\alpha_{\mathrm{m}}$, which is almost equal to that for single solitons; it also rapidly decreases with $d$. In contrast to single solitons, for bound states the rescaled energy $U_r$ is a monotonically increasing function of $\alpha$. When $\alpha \to \alpha_{\mathrm{m}}$ the energy flow $U_r$ diverges. Calculations suggest that stable extended trains of gray solitons are possible. Numerical simulations for bound states of two and three gray solitons showed that such states are completely stable at small and moderate velocities $\alpha$ (Figs. 5(c) and 5(d)), but may exhibit weak instabilities when $\alpha \to \alpha_\mathrm{m}$.

We thus conclude by stressing that a nonlocal nonlinearity drastically affect the profiles, velocities, and interactions between gray solitons. Particularly important is the finding of a maximum possible soliton velocity in nonlocal media that decreases fast with high nonlocality. Our findings suggest a feasible way to steer gray soliton light beams, by tuning the strength of nonlocality in different nonlinear media.

# References with titles

1. W. Krolikowski, O. Bang, N. I. Nikolov, D. Neshev, J. Wyller, J. J. Rasmussen, and D. Edmundson, "Modulational instability, solitons and beam propagation in spatially nonlocal nonlinear media," J. Opt. B **6**, S288 (2004).

2. M. Peccianti, K. A. Brzdakiewicz, and G. Assanto, "Nonlocal spatial soliton interactions in nematic liquid crystals," Opt. Lett. **27**, 1460 (2002).

3. G. Assanto and M. Peccianti, "Spatial solitons in nematic liquid crystals," IEEE J. Quantum Electronics **39**, 13 (2003).

4. X. Hutsebaut, C. Cambournac, M. Haelterman, A. Adamski, and K. Neyts, "Single-component higher-order mode solitons in liquid crystals," Opt. Commun. **233**, 211 (2004).

5. Z. Xu, Y. V. Kartashov, and L. Torner, "Upper threshold for stability of multipole-mode solitons in nonlocal nonlinear media," Opt. Lett. **30**, 3171 (2005).

6. Z. Xu, Y. V. Kartashov, and L. Torner, "Stabilization of vector soliton complexes in nonlocal nonlinear media," Phys. Rev. E **73**, 055601(R) (2006).

7. A. V. Mamaev, A. A. Zozulya, V. K. Mezentsev, D. Z. Anderson, and M. Saffman, "Bound dipole solitary solutions in anisotropic nonlocal self-focusing media," Phys. Rev. A **56**, R 1110 (1997).

8. S. Lopez-Aguayo, A. S. Desyatnikov, Y. S. Kivshar, S. Skupin, W. Krolikowski, and O. Bang, "Stable rotating dipole solitons in nonlocal optical media," Opt. Lett. **31**, 1100 (2005).

9. Y. V. Kartashov, L. Torner, V. A. Vysloukh, and D. Mihalache, "Multipole vector solitons in nonlocal nonlinear media," Opt. Lett. **31**, 1483 (2006).

10. S. Skupin, O. Bang, D. Edmundson, and W. Krolikowski, "Stability of two-dimensional spatial solitons in nonlocal nonlinear media," Phys. Rev. E **73**, 066603 (2006).

11. A. I. Yakimenko, V. M. Lashkin, and O. O. Prikhodko, "Dynamics of two-dimensional coherent structures in nonlocal nonlinear media," Phys. Rev. E **73**, 066605 (2006).

12. S. Lopez-Aguayo, A. S. Desyatnikov, and Y. S. Kivshar, "Azimuthons in nonlocal nonlinear media," Opt. Express **14**, 7903 (2006).
13. C. Rotschild, M. Segev, Z. Xu, Y. V. Kartashov, L. Torner, and O. Cohen, "Two-dimensional multi-pole solitons in nonlocal nonlinear media," Opt. Lett. **31**, 3312 (2006).
14. D. Briedis, D. Petersen, D. Edmundson, W. Krolikowski, and O. Bang, "Ring vortex solitons in nonlocal nonlinear media," Opt. Express **13**, 435 (2005).
15. A. I. Yakimenko, Y. A. Zaliznyak, and Y. S. Kivshar, "Stable vortex solitons in nonlocal self-focusing nonlinear media," Phys. Rev. E **71**, 065603(R) (2005).
16. C. Rotschild, O. Cohen, O. Manela, M. Segev, and T. Carmon, "Solitons in nonlinear media with an infinite range of nonlocality: first observation of coherent elliptic solitons and of vortex-ring solitons," Phys. Rev. Lett. **95**, 213904 (2005).
17. N. I. Nikolov, D. Neshev, W. Krolikowski, O. Bang, J. J. Rasmussen, and P. L. Christiansen, "Attraction of nonlocal dark optical solitons," Opt. Lett. **29**, 286 (2004).
18. A. Dreischuh, D. N. Neshev, D. E. Petersen, O. Bang, and W. Krolikowski, "Observation of attraction between dark solitons," Phys. Rev. Lett. **96**, 043901 (2006).
19. Y. S. Kivshar and B. Luther-Davies, "Dark optical solitons: physics and applications," Phys. Rep. **298**, 81 (1998).
20. A. Fratalocchi, G. Assanto, K. A. Brzdakiewicz, and M. A. Karpierz, "Optically induced Zener tunneling in one-dimensional lattices," Opt. Lett. **31**, 790 (2006).
21. W. J. Tomlinson, R. J. Hawkins, A. M. Weiner, J. P. Heritage, and R. N. Thurston, "Dark optical solitons with finite-width background pulses," J. Opt. Soc. Am. B **6**, 329 (1989).
22. D. Foursa and P. Emplit, "Investigation of black-gray soliton interaction," Phys. Rev. Lett. **77**, 4011 (1996).

# References without titles

1. W. Krolikowski, O. Bang, N. I. Nikolov, D. Neshev, J. Wyller, J. J. Rasmussen, and D. Edmundson, J. Opt. B **6**, S288 (2004).
2. M. Peccianti, K. A. Brzdakiewicz, and G. Assanto, Opt. Lett. **27**, 1460 (2002).
3. G. Assanto and M. Peccianti, IEEE J. Quantum Electronics **39**, 13 (2003).
4. X. Hutsebaut, C. Cambournac, M. Haelterman, A. Adamski, and K. Neyts, Opt. Commun. **233**, 211 (2004).
5. Z. Xu, Y. V. Kartashov, and L. Torner, Opt. Lett. **30**, 3171 (2005).
6. Z. Xu, Y. V. Kartashov, and L. Torner, Phys. Rev. E **73**, 055601(R) (2006).
7. A. V. Mamaev, A. A. Zozulya, V. K. Mezentsev, D. Z. Anderson, and M. Saffman, Phys. Rev. A **56**, R1110 (1997).
8. S. Lopez-Aguayo, A. S. Desyatnikov, Y. S. Kivshar, S. Skupin, W. Krolikowski, and O. Bang, Opt. Lett. **31**, 1100 (2005).
9. Y. V. Kartashov, L. Torner, V. A. Vysloukh, and D. Mihalache, Opt. Lett. **31**, 1483 (2006).
10. S. Skupin, O. Bang, D. Edmundson, and W. Krolikowski, Phys. Rev. E **73**, 066603 (2006).
11. A. I. Yakimenko, V. M. Lashkin, and O. O. Prikhodko, Phys. Rev. E **73**, 066605 (2006).
12. S. Lopez-Aguayo, A. S. Desyatnikov, and Y. S. Kivshar, Opt. Express **14**, 7903 (2006).
13. C. Rotschild, M. Segev, Z. Xu, Y. V. Kartashov, L. Torner, and O. Cohen, Opt. Lett. **31**, 3312 (2006).
14. D. Briedis, D. Petersen, D. Edmundson, W. Krolikowski, and O. Bang, Opt. Express **13**, 435 (2005).
15. A. I. Yakimenko, Y. A. Zaliznyak, and Y. S. Kivshar, Phys. Rev. E **71**, 065603(R) (2005).
16. C. Rotschild, O. Cohen, O. Manela, M. Segev, and T. Carmon, Phys. Rev. Lett. **95**, 213904 (2005).

17. N. I. Nikolov, D. Neshev, W. Krolikowski, O. Bang, J. J. Rasmussen, and P. L. Christiansen, Opt. Lett. **29**, 286 (2004).
18. A. Dreischuh, D. N. Neshev, D. E. Petersen, O. Bang, and W. Krolikowski, Phys. Rev. Lett. **96**, 043901 (2006).
19. Y. S. Kivshar and B. Luther-Davies, Phys. Rep. **298**, 81 (1998).
20. A. Fratalocchi, G. Assanto, K. A. Brzdakiewicz, and M. A. Karpierz, Opt. Lett. **31**, 790 (2006).
21. W. J. Tomlinson, R. J. Hawkins, A. M. Weiner, J. P. Heritage, and R. N. Thurston, J. Opt. Soc. Am. B **6**, 329 (1989).
22. D. Foursa and P. Emplit, Phys. Rev. Lett. **77**, 4011 (1996).

# Figure captions

Figure 1 (color online). Intensity and refractive index distributions for single gray solitons for $\alpha = 0.3$ (a) and $\alpha = 0.61$ (b), and for a bound state of two gray solitons for $\alpha = 0.3$ (c) and $\alpha = 0.58$ (d). In all cases $d = 5$.

Figure 2. Phase distributions for single gray solitons depicted in Figs. 1(a) and 1(b).

Figure 3. (a) Renormalized energy flow of single gray solitons versus velocity for $d = 5$. (b) Comparison of grayness versus velocity in local and in nonlocal media. Points marked by circles in (a) and (b) correspond to the solitons depicted in Figs. 1(a) and 1(b).

Figure 4. (a) Maximal velocity of gray soliton versus nonlocality degree. (b) Grayness versus nonlocality degree for $\alpha = 0.5$.

Figure 5. Stable propagation of perturbed gray solitons in nonlocal nonlinear medium with $d = 5$. (a) Single soliton with $\alpha = 0.3$, (b) single soliton with $\alpha = 0.61$, (c) bound state of two gray solitons with $\alpha = 0.3$, (d) bound state of three gray solitons with $\alpha = 0.3$.

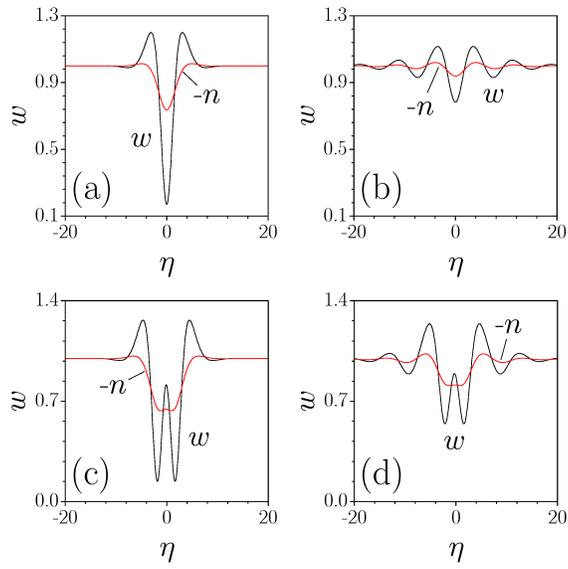

Figure 1 (color online). Intensity and refractive index distributions for single gray solitons for $\alpha = 0.3$ (a) and $\alpha = 0.61$ (b), and for a bound state of two gray solitons for $\alpha = 0.3$ (c) and $\alpha = 0.58$ (d). In all cases $d = 5$.

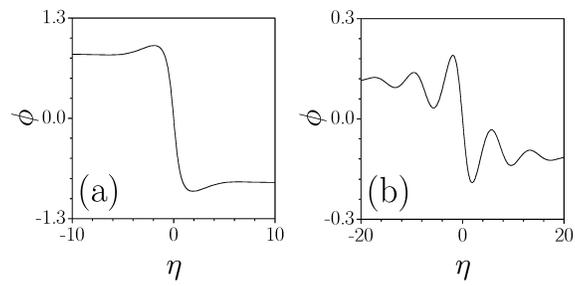

Figure 2.   Phase distributions for single gray solitons depicted in Figs. 1(a) and 1(b).

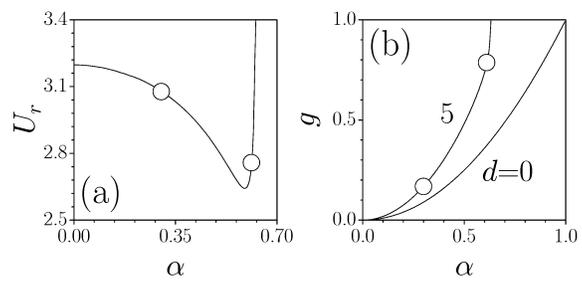

Figure 3. (a) Renormalized energy flow of single gray solitons versus velocity for $d = 5$. (b) Comparison of grayness versus velocity in local and in nonlocal media. Points marked by circles in (a) and (b) correspond to the solitons depicted in Figs. 1(a) and 1(b).

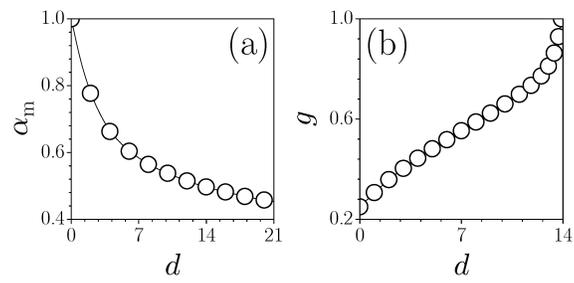

Figure 4. (a) Maximal velocity of gray soliton versus nonlocality degree. (b) Grayness versus nonlocality degree for $\alpha = 0.5$.

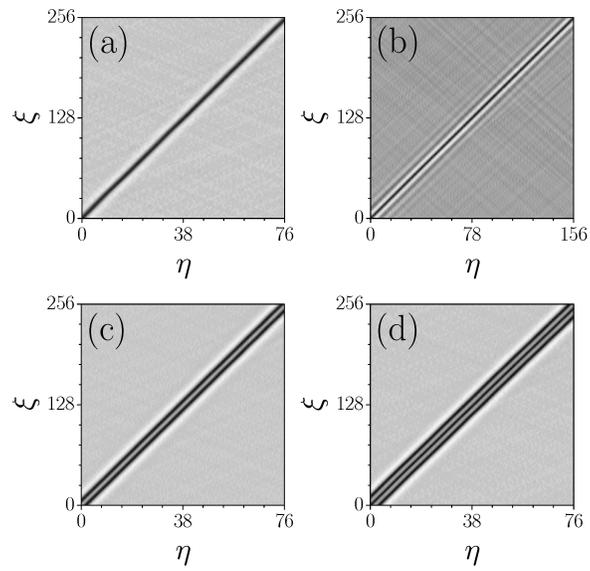

Figure 5. Stable propagation of perturbed gray solitons in nonlocal nonlinear medium with $d = 5$. (a) Single soliton with $\alpha = 0.3$, (b) single soliton with $\alpha = 0.61$, (c) bound state of two gray solitons with $\alpha = 0.3$, (d) bound state of three gray solitons with $\alpha = 0.3$.